\documentclass[namedreferences]{solarphysics}

\usepackage[hyperref,optionalrh,showbiblabels]{spr-sola-addons} 
\usepackage{graphicx}        
\usepackage{amssymb}        
\usepackage{bm}
\usepackage{color}           
\usepackage{breakurl}        

\newcommand{\etal}{{\it et al.}}
\newcommand{\Fig}[1]{Figure~\ref{#1}}
\newcommand{\q}{\mathit q}

\chardef\us=`\_

\begin{document}

\begin{article}
\begin{opening}

\title{Study of Sunspot Penumbra to Umbra Area Ratio using Kodaikanal White-light Digitized Data }

\author[addressref={aff1},corref,email={bibhuti.kj@iiap.res.in}]{\inits{B. K.}\fnm{Bibhuti Kumar}~\lnm{Jha}\orcid{0000-0003-3191-4625}}
\author[addressref={aff2},corref,email={smandal.solar@gmail.com}]{\inits{S.}\fnm{Sudip}~\lnm{Mandal}\orcid{0000-0002-7762-5629}}
\author[addressref={aff1,aff3},corref,email={dipu@iiap.res.in}]{\inits{D.}\fnm{Dipankar}~\lnm{Banerjee}\orcid{0000-0003-4653-6823}}


\address[id=aff1]{Indian Institute of Astrophysics, Koramangala, Bangalore 560034, India}
\address[id=aff2]{Indian Institute of Astrophysics, Koramangala, Bangalore 560034, India (Currently at Max Planck Institute for Solar System Research, Goettingen, Germany)}
\address[id=aff3]{Center of Excellence in Space Sciences India, IISER Kolkata, Mohanpur 741246, West Bengal, India}

\runningauthor{B. K. Jha \etal}
\runningtitle{Penumbra to Umbra Area Ratio from Kodaikanal Observatory}

\begin{abstract}
We study the long-term behaviour of sunspot penumbra to umbra area ratio by analyzing the recently digitized Kodaikanal white-light data (1923-2011). We implement an automatic umbra extraction method and compute the ratio over eight solar cycles (Cycles 16-23). Although the average ratio doesn't show any variation with spot latitudes, cycle phases and strengths, it increases from 5.5 to 6 as the sunspot size increases from 100 $\mu$hem to 2000 $\mu$hem. Interestingly, our analysis also reveals that this ratio for smaller sunspots (area $<$ 100 $\mu$hem) does not have any long-term systematic trend which was earlier reported from the Royal Observatory, Greenwich (RGO) photographic results. To verify the same, we apply our automated extraction technique on Solar and Heliospheric Observatory (SOHO)/Michelson Doppler Imager (MDI) continuum images (1996-2010). Results from this data not only confirm our previous findings, but also show the robustness of our analysis method.
\end{abstract}
\end{opening}
\section{Introduction}
\label{s-intro}
Sunspots, the most prominent features on the solar photosphere, appear dark when observed in visible wavelengths. They also show periodic variations in their properties over an $\approx$11 years time-scale, generally referred as the ``solar cycle" \citep{Hathaway2015}. In fact, after the observations by \cite{1908ApJ....28..315H}, it became clear that sunspots are the locations of strong magnetic fields ($\approx$4 kG) which inhibit convection within them. Due to such suppression of energy, they appear as dark structures \citep{Solanki2003}. A closer inspection of sunspot images reveals that there are, actually, two different features within a spot: a darker (with respect to photospheric intensity) umbra surrounded by a lighter penumbra. This contrast in appearance is generally attributed to different strengths and orientations of the magnetic fields which are present in these two regions \citep{2003A&A...410..695M}. Hence, area measurements of umbra and penumbra carry these magnetic field information too. The other importance of these measurements come from their application in calculating the \textit{Photometric Solar Index} (PSI) values which quantize the decrement of \textit{Total Solar Irradiance} (TSI) due to the presence of a spot on solar disc \citep{1977soiv.conf...93F,Hudson1982}. Thus, a knowledge of long-term variations in the umbra and penumbra area will enhance our understanding of solar variability.

One of the earliest measurements of umbra and penumbra area values was reported by \cite{1933PASP...45...51N} who studied almost one thousand unipolar or preceding member of bipolar sunspots from Royal Observatory, Greenwich (RGO) between 1917 to 1920. The average ratio ($ \q $), between the area of penumbra to that of umbra, was quoted as $\approx$4.7 and it was also found to be independent of sunspot sizes. However, examining the diameters of umbra and penumbra of 53 sunspots as photographed by Wolfer at Z\"urich, \cite{1939MiZur..14..439W} noted that the $\q$ value decreases from 6.8 to 3.4 as the sunspot area increases from $100~\mu$hem to $1000~\mu$hem. The first investigation of the long-term evolution of this ratio was reported by \cite{1955ApNr....5..167J, 1956AnAp...19..165J} where the authors analysed the RGO data from 1878 to 1945. Interestingly, they noted that the ratio is a decreasing function of sunspot size during cycle maxima but the variation is much lower than the values as reported in \cite{1939MiZur..14..439W}. Several follow up studies by \cite{1956ApNr....5..207T,1971BAICz..22..352A,1993SoPh..146...49B} also confirmed similar results by including more complex sunspots and larger statistics.

Using the largest set of observations as recorded in RGO data (161839 sunspot groups between 1874-1976), \cite{2013SoPh..286..347H} calculated the $\q $ values for each of these cases and noted that it increases from 5 to 6 as sunspot group size increases from $100~\mu$hem to $2000~\mu$hem. However, the author did not find any dependency of $\q $ on the cycle phase or the locations of the spots. The most remarkable result of all was the behaviour of smaller sunspot groups (area $<100~\mu$hem), for which the author found a substantial change in the $\q $ values within a relatively smaller timescale. The ratio decreased significantly from 7 to 3 during solar Cycles 14$-$16, however, it again increased to $>$7 in 1961 at the end of Cycle 19.

\section{Data}
\begin{figure}
\centering
\centerline{\includegraphics[width=\textwidth,clip=]{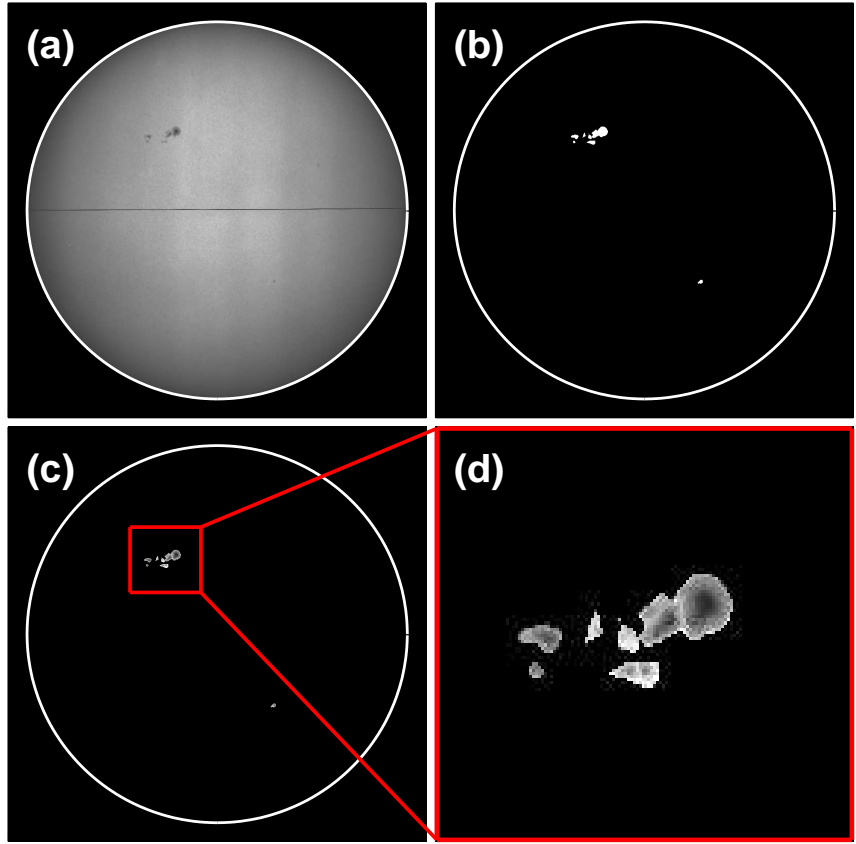}}
\caption{{\it Panel-(a):} A calibrated white-light image from Kodaikanal Observatory as recorded on 1955-01-07 08:15. {\it Panel-(b):} Binary image of the extracted sunspots. {\it Panel-(c):} Isolated spots in the original grey scale image produced by multiplying images on panel-(a) with panel-(b). A zoomed in view is presented in {\it Panel-(d).}}
\label{fig1}
\end{figure}

In this study, we have used the newly digitized and calibrated white-light full disk images (Figure~\ref{fig1}a) from Kodaikanal Solar Observatory. Details of this digitization, including the various steps of calibration process, are reported in \cite{2013A&A...550A..19R}. Recently, \cite{2017A&A...601A.106M} catalogued the whole-spot area series\footnote{This catalogue is available online at \url{https://kso.iiap.res.in/new/white_light}.} (between 1921 and 2011) by using a semi-automated sunspot detection algorithm on this data. We start our analysis with these detected binary images of sunspots as shown in Figure~\ref{fig1}b. In-order to isolate the spots, we multiply the binary mask with the limb-darkening corrected full disc images. The final results are displayed in Figure~\ref{fig1}c-\ref{fig1}d.

\section{Method}
Considering the volume of the data to be processed, we opted for an automatic boundary detection algorithm. A number of methods have already been used in the past to automatically detect umbrae of sunspots: \cite{1990SoPh..129..191B} \& \cite{2016JApA...37....3P} using fixed intensity threshold; \cite{1997SoPh..175..197P} using cumulative histogram method and \cite{1997joso.proc...89S} using the inflection method. Despite their successes on other data sets (mostly of smaller duration), we found that none of these methods actually produces a faithful result when applied on the entire Kodaikanal data. The main reasons behind this are the varying image quality over time, poor contrast, presence of artefacts {\it etc}. Keeping these limitations in mind, we select an adaptive umbra detection method based on the Otsu thresholding technique \citep{4310076}.
This method finds the optimum threshold for an image which has a bimodal intensity distribution. In our case, the two different intensity levels of umbra and penumbra constitute a similar type of distribution which is suitable for such an application. Mathematically, to calculate the threshold, this method maximizes the between-class variance of the distribution. If $t$ is the threshold that separates ${\rm L}$ bins of histogram in background class ($C_{\rm b}$) and foreground class ($C_{\rm f}$), then the probability of occurrence of background ($\omega _{\rm b}$) and foreground classes ($\omega _{\rm f}$) are
\begin{eqnarray}
   \omega _{\rm b}&=&\sum _{i=1}^{{\rm t}}P(i)=\omega(t),\\
   \omega _{\rm f}&=&\sum _{{i=t+1}}^{{\rm L}}P(i)=1-\omega(t)
\end{eqnarray}
where $P(i)$ represents the probability of occurrence of the $i_{\rm th}$ bin.
The between-class variance {\bf($\sigma_{\rm B}$)} of the distribution for a particular $t$ can be written as
\begin{equation}
 \sigma _{\rm B}(t)^2=\omega _{\rm b}(\mu _{\rm b}-\mu)^2+\omega _{\rm f}(\mu _{\rm f}-\mu)^2.
 \label{eq3}
\end{equation}
In Equation \ref{eq3}, $\mu$ (the mean of the distribution) and $\mu _{\rm b}$ and $\mu _{\rm f}$ (the means of the background and foreground class) are defined as
\begin{eqnarray}
    \mu&=&\sum _{i=1}^{{\rm L}}iP(i),\\
    \mu _{\rm b}&=&\sum _{i=1}^{t}iP(i/C_{\rm b})=\frac{\mu(t)}{\omega(t)},\\
    \mu _{\rm f}&=&\sum _{i=t+1}^{\rm L}iP(i/C_{\rm f})=\frac{\mu-\mu(t)}{1-\omega(t)}.
\end{eqnarray}
    
In this work, we use the {\texttt cgotsu$\_$threshold.pro}\footnote{Description is available at \url{http://www.idlcoyote.com/idldoc/cg/cgotsu_threshold.html}.} routine, an IDL\footnote{For more details, visit \url{ https://www.harrisgeospatial.com/Software-Technology/IDL}.} implementation of the above concept.

\begin{figure}
\centering
\centerline{\includegraphics[width=\textwidth,clip=]{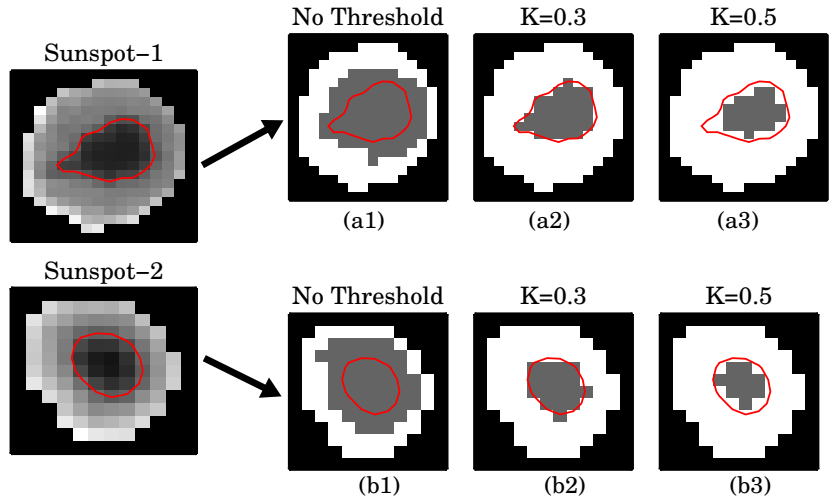}}
\caption{Two representative examples of our umbra detection technique on Kodaikanal sunspot data. The red contours (in panels (a1-b3)) highlight the umbra-penumbra boundaries as estimated by eye whereas the detected umbrae for different set of threshold values are shown as grey regions. See text for more details.}
\label{fig2}
\end{figure} 

 We demonstrate the application of this algorithm on our data with two representative examples as shown in Figure~\ref{fig2}. The red contours on the spots represent the umbra-penumbra boundary as estimated by visual inspections. We expect an umbral boundary, as detected by this Otsu method, to more-or-less coincide with this contour. When applied on the original image, the detected umbra comes out to be significantly larger in size as seen in panels (a1,b1) of Figure~\ref{fig2}. Upon investigation, we realize that this over-estimation occurs due to the presence of few brighter pixels on the edge of the detected spots. In fact, these bright pixels are originally a part of the quiet Sun region and got picked up during the sunspot detection procedure. Though the number of such pixels is very small compared to the total pixels of a typical sunspot, it seems to have a significant influence on the derived threshold value. To get rid of these ``rouge pixels", a pre-processing technique is applied before feeding the spots into the Otsu method. We set up an intensity filter which is based on a threshold defined as:

\begin{equation}
I_{\rm th}=\bar{I}-k\sigma
\label{eq2}
\end{equation}
where $\bar{I}$ and $\sigma$ are mean and standard deviation of spot region. With this criteria, a pixel with intensity ($ I_{\rm n}$) greater than $I_{\rm th}$ gets removed form that specific spot {\it i.e.} we set $I_{\rm n}=0$. Although, from Equation~\ref{eq2}, we note that {\it k} is a free parameter which needs to be optimized. We fix this issue by taking a large subset of randomly chosen sunspots (of different contrasts and morphologies) and repeating the above procedure with multiple values of {\it k}. After visual inspections of each of those results, we find that two values, $ k=0.3$ and $k=0.5$, produce the most accurate results as compared to other {\it k} values. However, most often or not, the umbra gets underestimated with $k = 0.5$ (Figure~\ref{fig2}a3, \ref{fig2}b3).

 \begin{figure}
\centering
\centerline{\includegraphics[width=\textwidth,clip=]{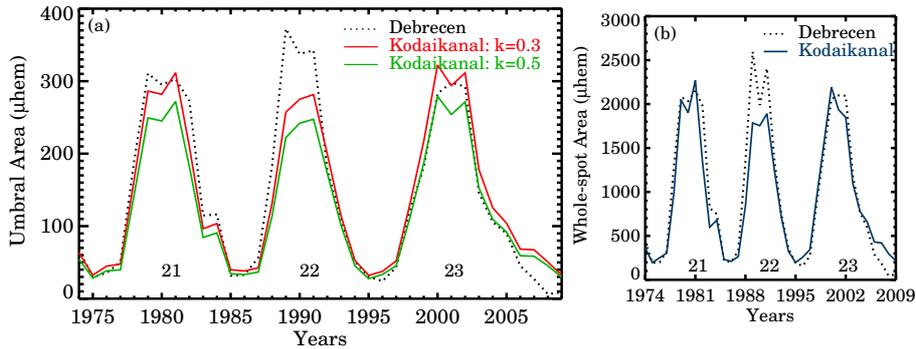}}
\caption{{\it Panel(a):} Comparison of  yearly averaged umbral areas between Kodaikanal ($k=0.3$ (red) \& $k=0.5$ (green)) and Debrecen data (black). {\it Panel(b):} Same as before but for the whole-spot area.}
\label{fig3}
\end{figure}
To better visualize this effect, we compare our results with the umbra measurements from Debrecen Observatory\footnote{This data is downloaded from \url{http://fenyi.solarobs.csfk.mta.hu/ftp/pub/DPD/data/dailyDPD1974_2016.txt}.} \citep{2016SoPh..291.3081B, 10.1093/mnras/stw2667} as shown in Figure~\ref{fig3}a. The plot highlights the fact that $k=0.3$ is indeed a better choice for our Kodaikanal data. However, there is a large discrepancy between the Kodaikanal values with that from Debrecen, near the Cycle 22 maxima. To eliminate the possibility of this being an artefact of our umbra detection technique, we also plot the whole spot area between the two observatories in Figure~\ref{fig3}b. Presence of a similar difference in this case too, indicates an underestimation of total sunspot area during the original spot detection procedure, as reported in \cite{2017A&A...601A.106M}.

Finally, we compute the penumbra to umbra area ratio as:
\begin{equation}
{\rm Ratio }= \q = \frac{A_{\rm W}}{A_{\rm U}}-1
\end{equation}
where $A_{\rm W}$ and $A_{\rm U}$ are whole spot area and umbra area. This definition is same as \cite{1971BAICz..22..352A} and \cite{2013SoPh..286..347H}. 

\section{Results}
We calculate the the ratio  $\q$ for the whole period of the currently available Kodaikanal data which covers Cycle 16 to Cycle 23. Different aspects of this ratio are discussed in this section.

\begin{figure}
\centerline{\includegraphics[width=\textwidth,clip=]{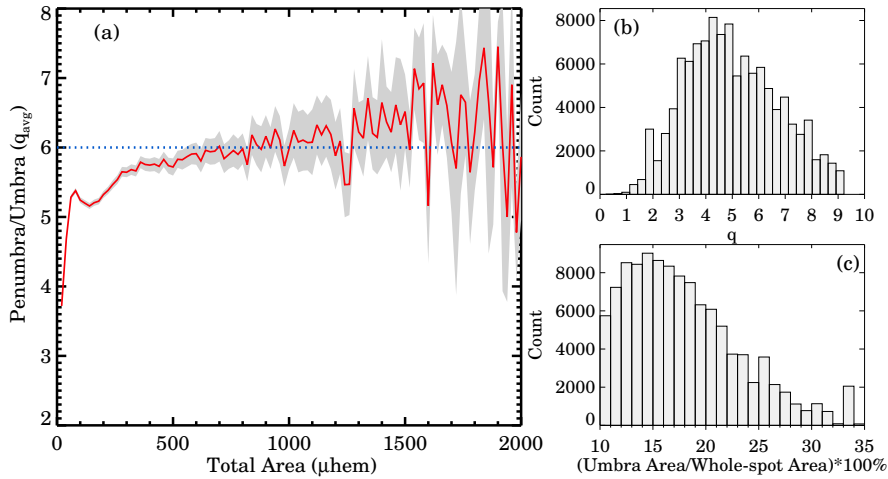}}
\caption{ {\it Panel(a):} Penumbra to umbra area ratio as function of total sunspot area binned over $20~\mu$hem. Grey shaded region represent the $2\sigma$ errors. {\it Panel(b)} shows the distribution of individual ratio ($\q$) whereas the distribution of percentage coverage of umbral area over the whole-spot area is shown in {\it Panel(c)}.}
\label{fig4}
\end{figure}

\subsection{Individual Variations}

To investigate the overall behaviour of  $\q$, we group the sunspot areas into bin sizes of $20~\mu$hem between $20-2000~\mu$hem and calculate the average ratio ({$\q_{\rm avg}$}) for all the sunspots falling in that particular bin. \Fig{fig4}a shows the quantity {$ \q_{\rm avg}$} as a function of total spot area. The shaded region represents the standard error of $2\sigma$ uncertainty. The error bars beyond area $>1500~\mu$hem are considerably larger due to poor statistics in those bins. Initially, the ratio for smaller spots (area $<100~\mu$hem), increases rapidly from 3.4 to 5.2. As the area increases further ($>100~\mu$hem), {$ \q_{\rm avg}$} tends to settle down to a value of $\approx$6 \citep{jha_mandal_banerjee_2018}. In fact, these results are consistent with the findings by \cite{1971BAICz..22..352A} \& \cite{2013SoPh..286..347H}. Physically this means larger spots tend to have larger penumbra (the observed slow upward trend), however large uncertainties make this conclusion rather weak. In addition to this, we note that there is a local minima of {$\q_{\rm avg}$} around 150~$\mu$hem which also needs further investigation and we do not have a convincing explanation for the same. Behavior of $\q$ for every detected sunspots is also analysed and presented in a histogram as shown in \Fig{fig4}b. The distribution peaks $\approx$4.5 and falls rapidly on both sides from the peak. Another interesting aspect is the coverage of umbra with respect to the total area for any individual sunspot. \Fig{fig4}c shows the distribution of this quantity (expressed in \%). The distribution peaks at 15\%, although there are significant number of cases between 15\% to 25\%. These properties are in good  agreement with previously measured values by \cite{2011A&A...533A..14W,2018SoPh..293..104C}. 

\subsection{Dependency on Cycle Strength and its Phases}
During the onset of a solar cycle, we see very few spots present on the disc (mostly of smaller sizes \cite{2017A&A...601A.106M}). They are also located at higher latitudes and with the progress of the cycle, they move towards the equator to form the popular `sunspot butterfly diagram'. 

\begin{figure}
\centerline{\includegraphics[width=\textwidth,clip=]{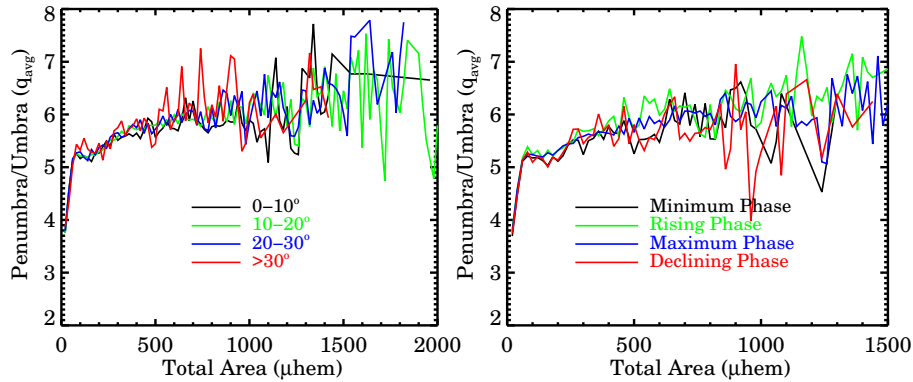}}
\caption{{\it Panel (a):} Variation of $\q_{\rm avg}$ as a function of total area in four different latitude bands as written on the panel; {\it Panel (b):} Same as previous but separated for four different activity phases of cycles.}
\label{fig5}
\end{figure}
We look for any such dependency of $\q_{\rm avg}$ by dividing the solar disc into several latitudinal bands. We fold the two hemispheres together and the results are plotted in \Fig{fig5}a. As seen from the plot, we find that the ratio does not depend on the latitude of a spot \citep{1971BAICz..22..352A,2013SoPh..286..347H}. In a slightly different representation of the same phenomena, we isolate the spots according to their appearances during a solar cycle. In fact, we are also motivated by some of the earlier studies by \cite{1955ApNr....5..167J,1956ApNr....5..207T,1971BAICz..22..352A}, where these authors reported different values of $\q_{\rm avg}$ during a cycle maxima as opposed to a cycle minima. To check this, a cycle is divided into four phases: minimum phase, rising phase, maximum phase and declining phase. The definition of each of these phases is the same as described in \cite{2013SoPh..286..347H}. Considering all the cycles together, we generate a plot as shown in \Fig{fig5}b. In this case, too, we do not notice any change for a given spot range in different phases of cycles. This is consistent with the RGO data as found by \cite{2013SoPh..286..347H}.
\begin{figure}
\centerline{\includegraphics[width=\textwidth,clip=]{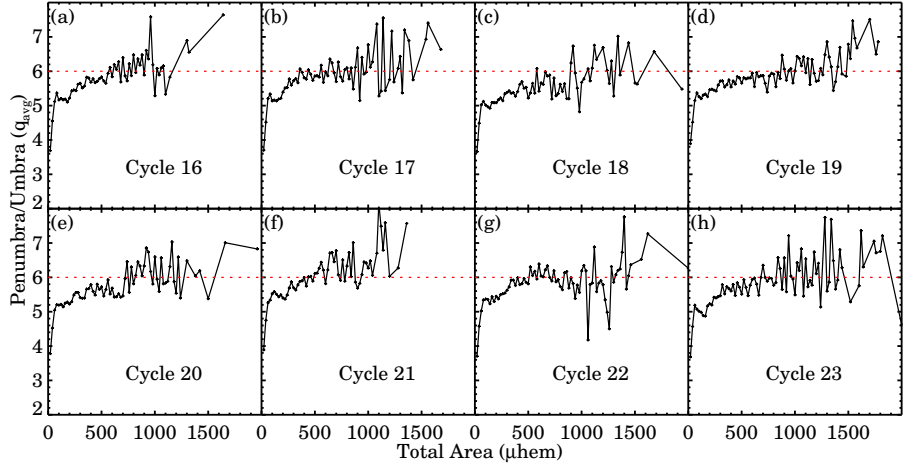}}
\caption{Figure above (a-h) shows the variations in $\q_{\rm avg}$ as recorded for each solar cycle (Cycles 16-23). Dashed red line is plotted just for reference.}
\label{fig6}
\end{figure}

 The other factor to potentially affect this ratio is the strength of a cycle. Similar spots in a weak cycle (Cycle 16) may have different $\q_{\rm avg}$ values than a strong cycle (Cycle 19). From Figure~\ref{fig6}a-\ref{fig6}h we note that there is absolutely no variation of $\q_{\rm avg}$ with cycles of different strengths. In a similar analysis by \cite{2013SoPh..286..347H} on RGO data, showed two different behaviours, specifically for the smaller spots (area $<100~\mu$hem), between even and odd numbered cycles. However, we do not find any such relation in our data.

\begin{figure}
\centering
\centerline{\includegraphics[width=\textwidth,clip=]{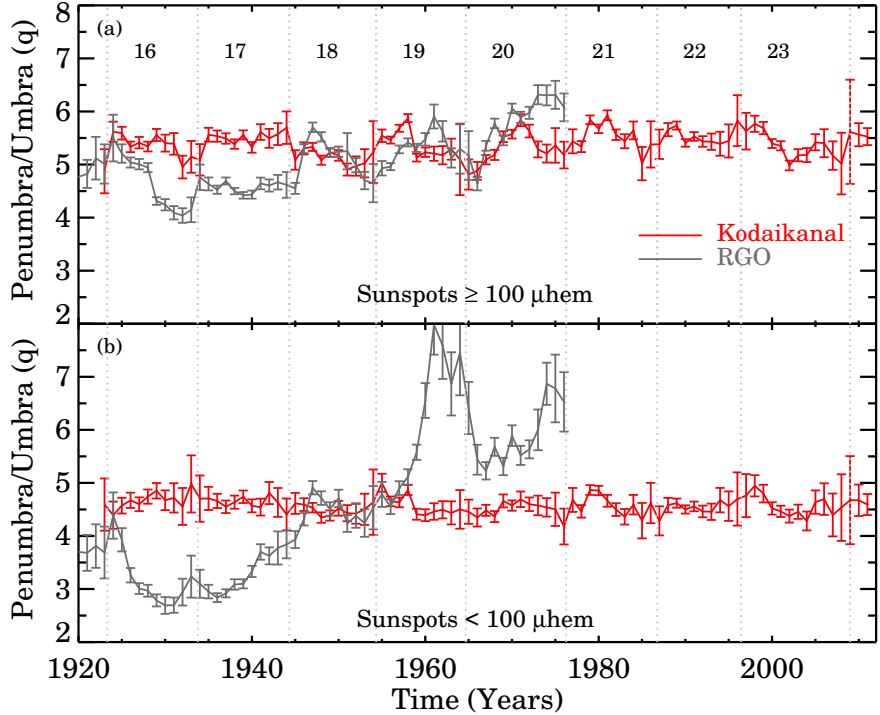}}
\caption{ Yearly averaged values of $\q$ as obtained from Kodaikanal data (red points) for two sunspot classes; for area $\geq$100~$\mu$hem ({\it Panel-(a)}) and for area $<$100~$\mu$hem ({\it Panel-(b)}). Similar values from RGO are also over plotted (grey points) for comparison. Error bars in each case represents the 2$\sigma$ uncertainties.}
\label{fig7}
\end{figure}

\subsection{Behaviour of Smaller and Larger Spots}

Sunspots of different sizes tend to show different behaviour \citep{2016ApJ...830L..33M}. In this section, we look for the temporal evolution of $\q $ from two class of sunspots: {\it i)} Sunspots with area $<$100~$\mu$hem (\Fig{fig7}a); {\it ii)} Sunspots with area $>$100$~\mu$hem (\Fig{fig7}b). The choice of this threshold at 100 $\mu$hem is primarily dictated by the fact that we see a jump in $\q$ value at this area value in \Fig{fig4}a . In order to compare our results with \cite{2013SoPh..286..347H}, we over plot the $\q$ values for RGO data as shown in \Fig{fig7}. For spots $>$100$~\mu$hem, the ratio neither show any significant time variation, nor any tendency to follow the solar cycles. The over plotted RGO data is in accordance with our values, except some systematically lower values during Cycles 16 to 17. One of the highlights of the work by \cite{2013SoPh..286..347H}, was the large secular variation of the ratio for smaller spots which showed 300\% increment with time. However, this property is not visible from Kodaikanal data which shows the ratio remains constant at $\approx$4.5 throughout the duration. In fact, analysing the Coimbra Astronomical Observatory (COI) data, \cite{2018SoPh..293..104C} also reported the absence of any type of secular variation in smaller spots.

As mentioned in the introduction, differences in the derived $\q$ values largely depend on the methods that have been used to detect umbra-penumbra boundary \citep{Steinegger1997}. Our method of Otsu thresholding has not been utilized in the literature before and thus, we feel the need of checking the robustness of this method on other independent datasets. The following section describes the application of the same on the space-based SOHO/MDI continuum images.

\subsection{Application on SOHO/MDI}
\begin{figure}
\centering
\centerline{\includegraphics[width=\textwidth,clip=]{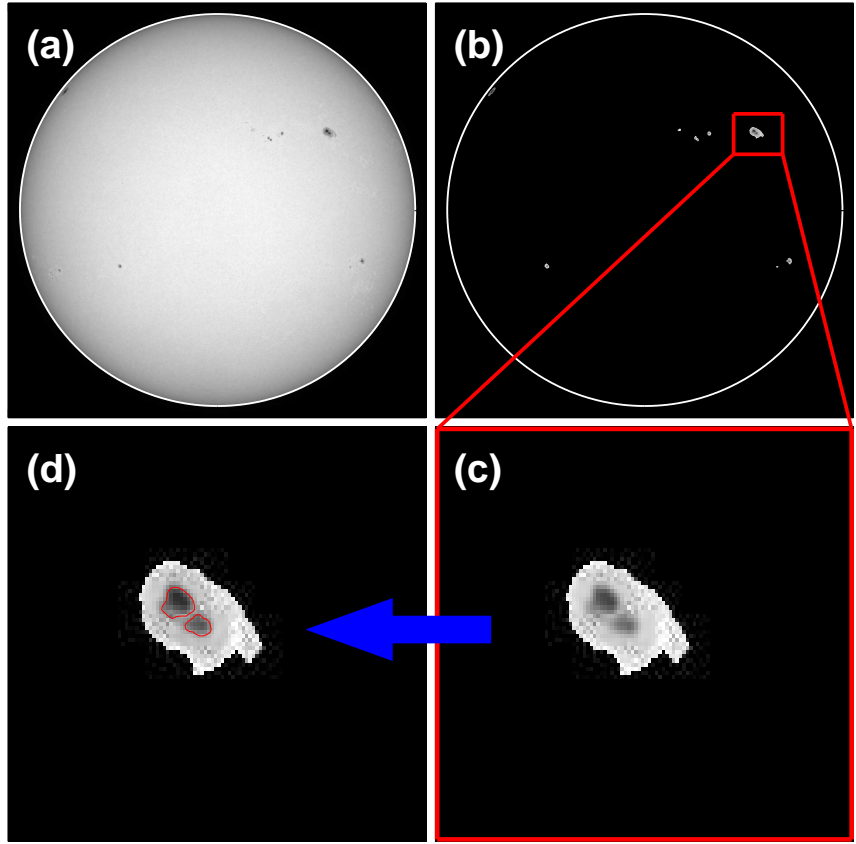}}
\caption{ Detection of umbra from SOHO/MDI data. {\it Panel-(a):} A representative continuum image as captured on 1999-05-14 23:59. {\it Panel-(b)} and {\t (c):} Detected sunspots and its zoomed in view respectively. {\it Panel-(d):} Contours of the umbrae over plotted onto the spot.}
\label{fig8}
\end{figure}

We analyse SOHO/MDI \citep{1995SoPh..162..129S} continuum images from 1996 to 2010 with a frequency of one image per day. First, we detect the sunspots using the same Sunspot Tracking and Recognition Algorithm (STARA: \cite{Watson2009}) as used on the Kodaikanal data. The detected spots are then fed to the Otsu algorithm for umbra detection. \Fig{fig8} summarizes the whole procedure.

\begin{figure}
\centering
\centerline{\includegraphics[width=\textwidth,clip=]{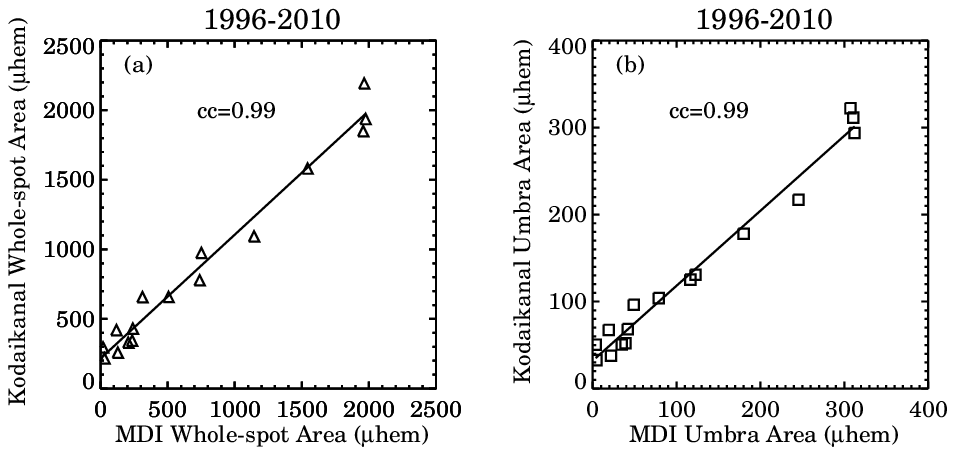}}
\caption{Comparison of yearly averaged whole sunspot area and umbra area as extracted from MDI and Kodaikanal.}
\label{fig9}
\end{figure}

We first compare the whole spot area values between Kodaikanal and MDI and the result is shown in \Fig{fig9}a. Computed yearly averages of whole spot areas are very similar to each other ({\rm c.c}=0.99). A similar behaviour is found for the umbral areas too (\Fig{fig9}b). Hence, the overall spot areas measured from these two observatories, show similar trends. However, our prime interest in this case, is to recover the behaviour of small ($<$100 $\mu$hem) spots as seen from Kodaikanal. 

\begin{figure}
\centering
\centerline{\includegraphics[width=\textwidth,clip=]{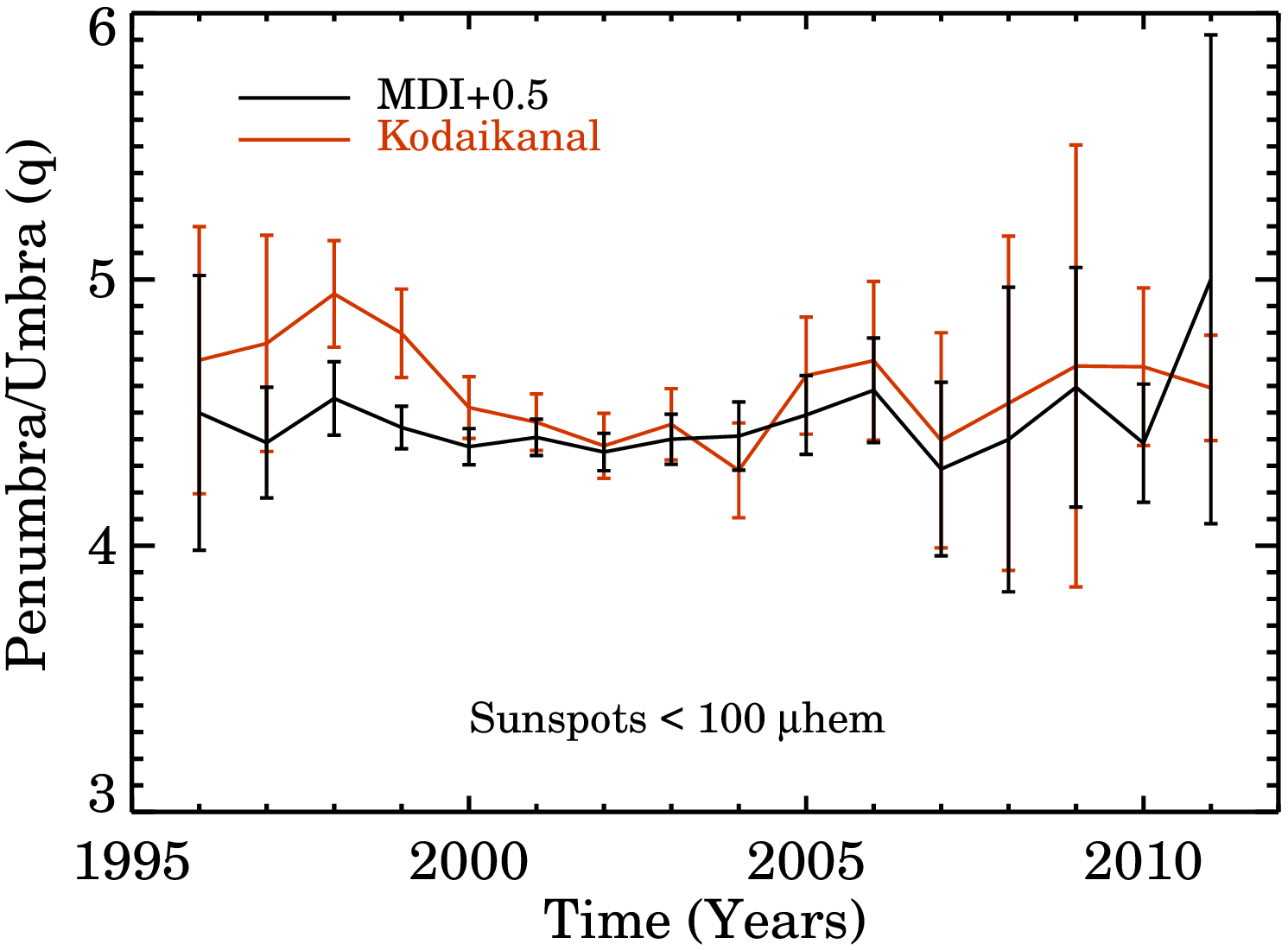}}
\caption{Ratio of areas of penumbra to umbra as a function of time for smaller sunspots.}
\label{fig10}
\end{figure}

In \Fig{fig10} we plot the $\q$ values (black solid line) for spots with area $<$100$\mu$hem as calculated from MDI. Kodaikanal values, for the overlapping period, are also over plotted (in red) for the ease of comparison. We see similar trends in both the curves, however the MDI values are needed to scale up by adding a constant factor of 0.5 to match the absolute values of Kodaikanal. This underestimation in MDI data is again, primarily due to the bright pixels present near the spot boundaries. During our analysis, we learnt that it is impossible to completely avoid these bright pixels while using the Sunspot Tracking and Recognition Algorithm (STARA) on large datasets. We can get around this problem by using a suitable {$k$} value as used in the earlier case. However, such a treatment only scales the absolute values, not the trend. Hence we present the results as it is.

\section{Conclusion}

In this paper, we investigated the long-term evolution of sunspot penumbra to umbra area ratio primarily using Kodaikanal white-light data. The main findings are summarized below:

$\bullet$ A total of 8 solar cycle (Cycles 16-23) data of Kodaikanal white-light digitized archive (1923-2011) and 15 years of MDI data (1996-2010) have been analysed in this work. We have used an automated umbra detection technique based on Otsu thresholding method and found that this method is efficient in isolating the umbra from variety of spots with different intensity contrasts.

$\bullet$ The penumbra to umbra ratio is found to be in the range of 5.5 to 6 for the spot range of 100~$\mu$hem to 2000~$\mu$hem. It is also found to be independent of cycle strengths, latitude zones and cycle phases. These results are in agreement with the previous reports in the literature.

$\bullet$ We segregated the spots according to their sizes and found that there is no signature of long-term secular variations for spots $<$100~$\mu$hem. This results contradicts the observations made by \cite{2013SoPh..286..347H} using the RGO data. However, our results are in close agreement with a recent study by \cite{2018SoPh..293..104C}.

$\bullet$ To check the robustness of our umbra detection technique, we analysed SOHO/MDI continuum images. These results also confirmed our previous findings from Kodaikanal data including the absence of any trend for smaller spots. During this study, we realized that although the Otsu technique is robust and adaptive in determining the umbral boundaries, it is also sensitive to any presence of artefacts within the spots.

In future, we plan to continue our study using the Solar Dynamics Observatory (SDO)/Helioseismic and Magnetic Imager (HMI) \citep{2012SoPh..275..229S} data. This will not only extend the time series but will also allow us to study the effect of higher spatial resolution ({\it i.e.} more pixels within a spot) in determining the optimum threshold. We also plan to use the Debrecen sunspot images (which are available online) and repeat the measurements of this ratio using our method. Debrecen has more than fifty years of overlap with Kodaikanal which makes this data suitable for cross calibration too. 
\begin{acks}
{Kodaikanal solar observatory is a facility of Indian Institute of Astrophysics, Bangalore, India. The data utilised in this article is now available for public use at http://kso.iiap.res.in. We would like to thank Ravindra B. and Manjunath Hegde for their tireless support during the digitization, calibration and sunspot detection processes. The authors also   acknowledge Subhamoy Chatterjee for his useful suggestions during the entire process.}
\end{acks}

\paragraph*{\footnotesize Disclosure of Potential Conflicts of Interest}
The authors declare that they have no conflicts of interest.
   

\bibliographystyle{spr-mp-sola}
\bibliography{umbra_penumbra_2019}  

\IfFileExists{\jobname.bbl}{} {\typeout{}
\typeout{****************************************************}
\typeout{****************************************************}
\typeout{** Please run "bibtex \jobname" to obtain} \typeout{**
the bibliography and then re-run LaTeX} \typeout{** twice to fix
the references !}
\typeout{****************************************************}
\typeout{****************************************************}
\typeout{}}



\end{article} 

\end{document}